\documentclass[]{article}
\usepackage[dvips,%
]{graphicx}
\begin{document}
\author{M.\,O.\,Gallyamov, I.\,V.\,Yaminsky}
\title{Visualization of atomic structure using AFM: theoretical description}
\date{
\emph{Physical department, Lomonosov Moscow State University, Moscow, Russia, 119899, \
e-mail.: glm@spm.phys.msu.su, tel.: +7(095)9392982, fax: +7(095)9392988
}
}
\maketitle
\begin{abstract}
We suggest simple model of image formation in atomic force microscope (AFM) taking into account contact deformations of probe and sample during scanning. The model explains the possibility of AFM visualization of regular atomic or molecular structure when probe-sample contact area is greater than an area per atom or molecule in surface lattice. (This is usually the case when lattice to be studied has subnanometer unit cell parameters and AFM investigations are carried out in air in contact mode). Two special peculiarities of AFM visualization of two-dimensional lattice could be observed under such conditions: 1) the inversion of contrast of AFM images, and 2) visualization of ``false atom'' under single atomic vacancy of surface studied. 
\end{abstract}

It is well known that AFM\,\cite{i_8} allows to visualize two-dimensional lattice of atoms or molecules packing and to determine unit cell parameters. It was proved in AFM studies of different objects: surfaces of covalent\,\cite{A432, A022, A489, K338}, ionic\,\cite{A304, A605}, molecular \,\cite{B272} including organic\,\cite{Cr185, FS015, B092} crystals, crystalline polymers\,\cite{P252}, biomembranes\,\cite{B171} and thin films of different nature (self-assembled\,\cite{W010}, Langmuir-Blodgett\,\cite{F183, A147} and similar films\,\cite{FS033}). The lattice parameters measured by AFM correlate usually well with data of diffraction techniques, so in this cases authors usually state that AFM provide subnanometer (``atomic'') resolution.

But there is a knowledge that in many cases AFM image of atomic or molecular lattice is not exactly ``true'' one\,\cite{A710, A531} due to extended tip-sample contact area. The main difference between ``true'' and ``false'' AFM images is that the latter does not allow to see individual atomic vacancies and in some cases does not represent adequately the unit cell. ``True'' AFM image could only be achieved with minimization of tip-sample interaction in very special conditions (in aqueous media\,\cite{B171, A531}, in ultrahigh vacuum using noncontact mode\,\cite{A569, K338}) when radius of contact area become smaller then unit cell parameters. Indeed when we carry out AFM experiment in contact mode under air conditions, it is possible to estimate tip-sample contact area using well known Herz theory\,\cite{landau7}:
\begin{equation}
\label{limit1}
a=(FDR)^{1/3}.
\end{equation}
Where
$$
D=\frac{3}{4}\left(\frac{1-\sigma^2}{E}+\frac{1-\sigma'^2}{E'}\right),
$$
and $E$, $E'$, $\sigma$, $\sigma'$ are Young and Poisson modules respectively.

Using eq.~(\ref{limit1}) we obtain 0.5--1.5\,nm values for radius of the contact area taking into account typical values of tip apex radius and applied force (minimized) in contact mode. In the case of studying of soft samples, for example organic thin films we obtain 2--4\,nm for the same parameter. The values are regarded sometimes as limit for spatial resolution in AFM\,\cite{A743}). 

We suppose that such non-point tip-sample interaction is the main reason of ``false'' AFM images formation. It is shown in\,\cite{A531}, that ``true'' AFM visualization are replaced by ``false'' one when applied force is increased during scanning (that means that contact area is increased also). In\,\cite{T177, T185, A710} "false" image formation was explained as a result of non-point nature of tip-sample interaction, but the authors regard very specific tips (with many atoms interacting with sample or with a flake on the apex). In this study we present a general theoretical model valid for a spherical approximation of tip apex. This model corresponds to the most common case in AFM and can be in principle experimentally checked.  

\subsection*{The model of AFM visualization}

\begin{figure}
\begin{center}
a)\includegraphics[width= 0.3\textwidth]{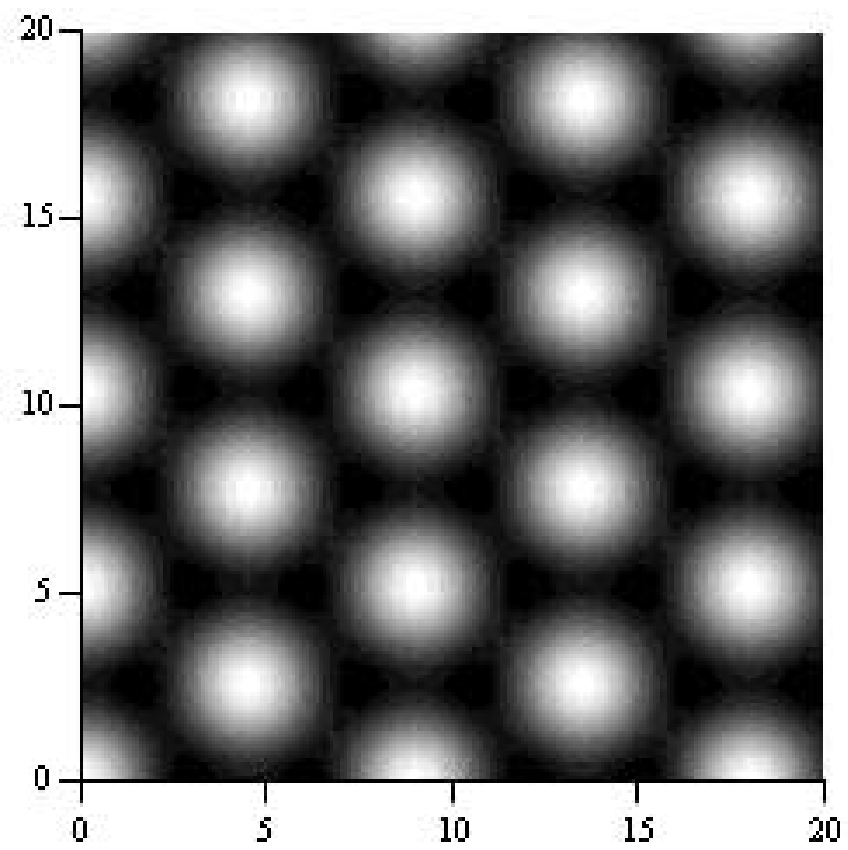}
b)\includegraphics[width= 0.3\textwidth]{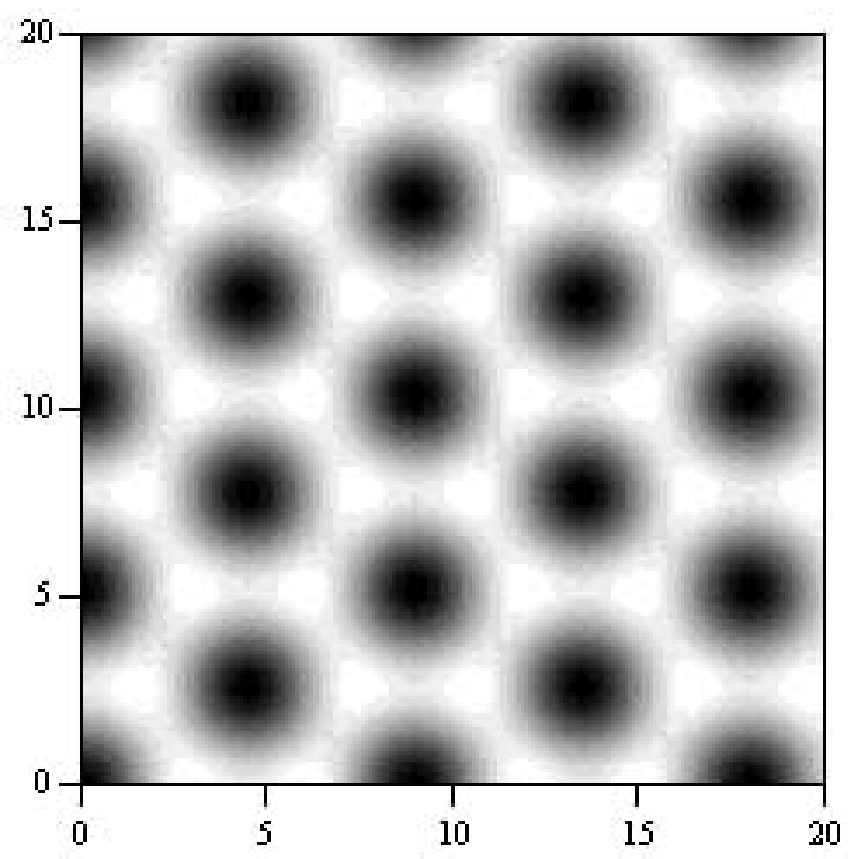}
c)\includegraphics[width= 0.3\textwidth]{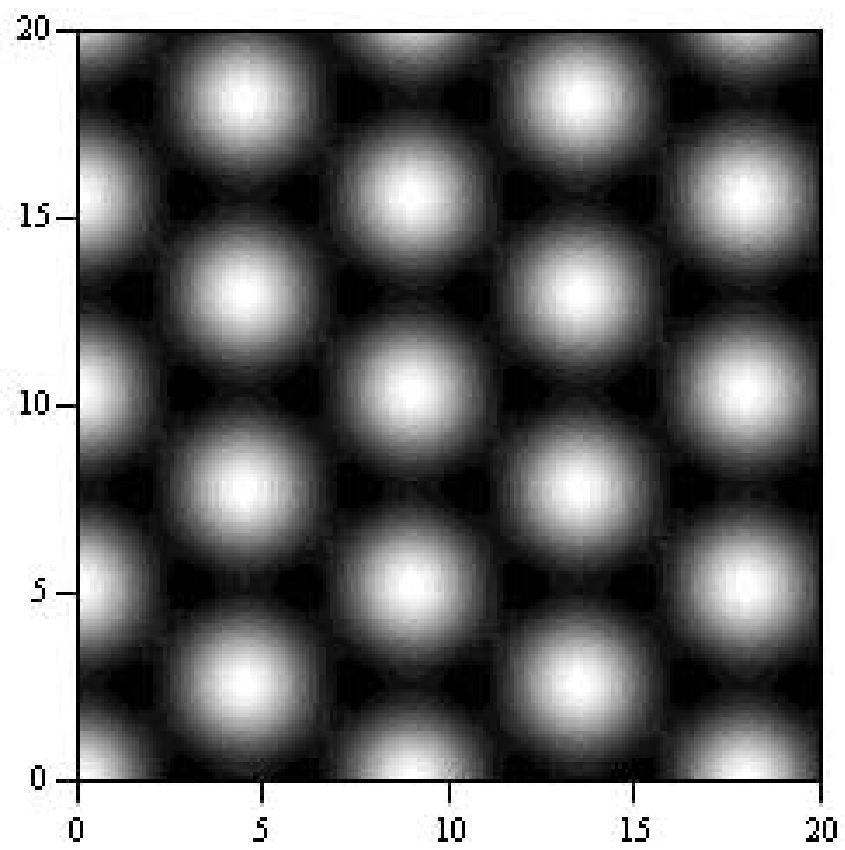}
d)\includegraphics[width= 0.3\textwidth]{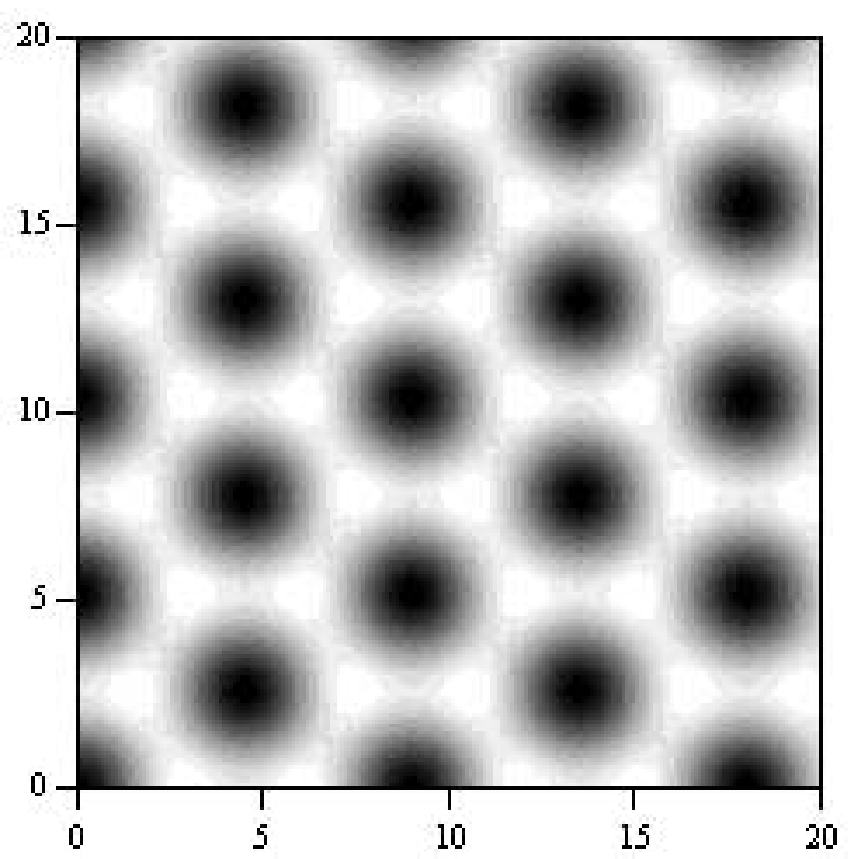}
e)\includegraphics[width= 0.3\textwidth]{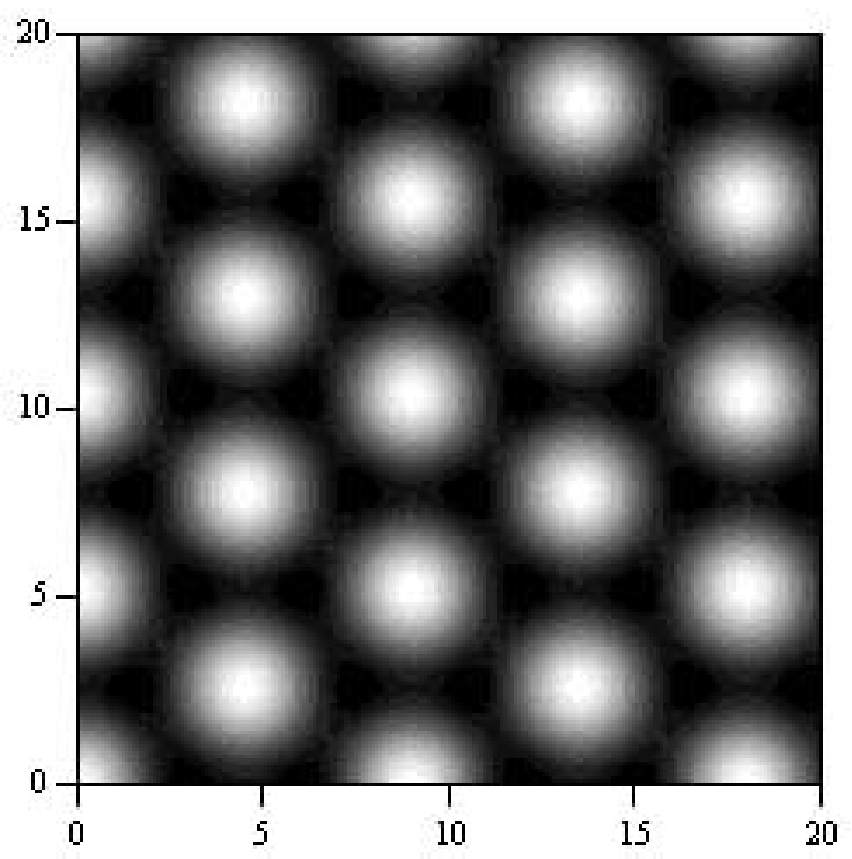}
f)\includegraphics[width= 0.3\textwidth]{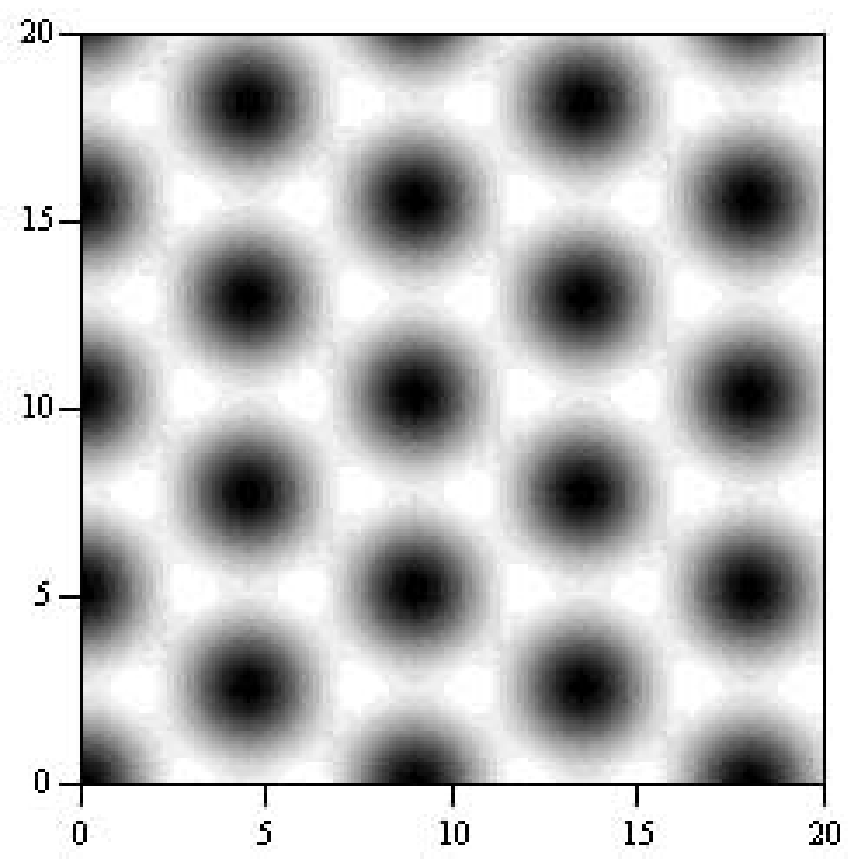}

\caption[]{Initial function~(\ref{phi}) (à) and AFM images calculated for different values of $a$ (0{,}9\,nm (b), 1{,}1\,nm (c), 1{,}3\,nm (d), 1{,}6\,nm (e) and 1{,}8\,nm (f)) using equation~(\ref{appar3}). \medskip \protect\\
{\footnotesize
The AFM images (b), (d) and (f) demonstrate inversion of contrast 
}
}
\label{three}
\end{center}
\end{figure}

Due to non uniform distribution of pressure in contact area\,\cite{landau7}:
$$
P(x,y)=\frac{3F}{2\pi a^2}\sqrt{1-\frac{x^2+y^2}{a^2}},
$$
(where $a$ is the radius of contact area, see. Eq\,\ref{limit1})
we could introduce model function for AFM visualization: 
\begin{equation}
\label{appar2}
\begin{array}{c}
\displaystyle
A(x-x',y-y')\simeq \\
\mbox{ }\\
\simeq
\left\{
\begin{array}{ll}
\displaystyle
\sqrt{1-\frac{(x-x')^2+(y-y')^2}{a^2}} &
\displaystyle
\mbox{if~} 
(x-x')^2+(y-y')^2 \leq a^2\\
\mbox{ }\\
\displaystyle
0 &
\displaystyle
\mbox{if~} (x-x')^2+(y-y')^2 > a^2,
\end{array}
\right. \\
\end{array}
\end{equation}
(The model is true for spherical tip apex and, consequently, correspond to a circular contact area. It is possible also to consider an elliptical contact area which correspond to the case of axial asymmetry of the probe.) The introduced function allows to give proper weight of each atom contribution in total image formation $f(x,y)$. 

\begin{equation}
\label{appar3}
f(x,y)=\int \!\! \int A(x-x',y-y')\varphi(x',y')\, dx'dy'
\end{equation} 
where $\varphi(x,y)$ presents atomic surface structure to be studied. We applied the model for the case of hexagonal packing of atoms with lattice parameter $d$.
\begin{equation}
\label{phi}
\varphi (x,y)= e^{ikx}+e^{ik(x + y\sqrt {3})/2}+ e^{ik(-x+y\sqrt {3})/2},
\end{equation}
where $k=4\pi /d \sqrt{3}$. The result of calculation for different values of $a$ and $d=5.2\,nm$ are presented on fig.\,\ref{three}.

\begin{figure}
\begin{center}
a)\includegraphics[width= 0.3\textwidth]{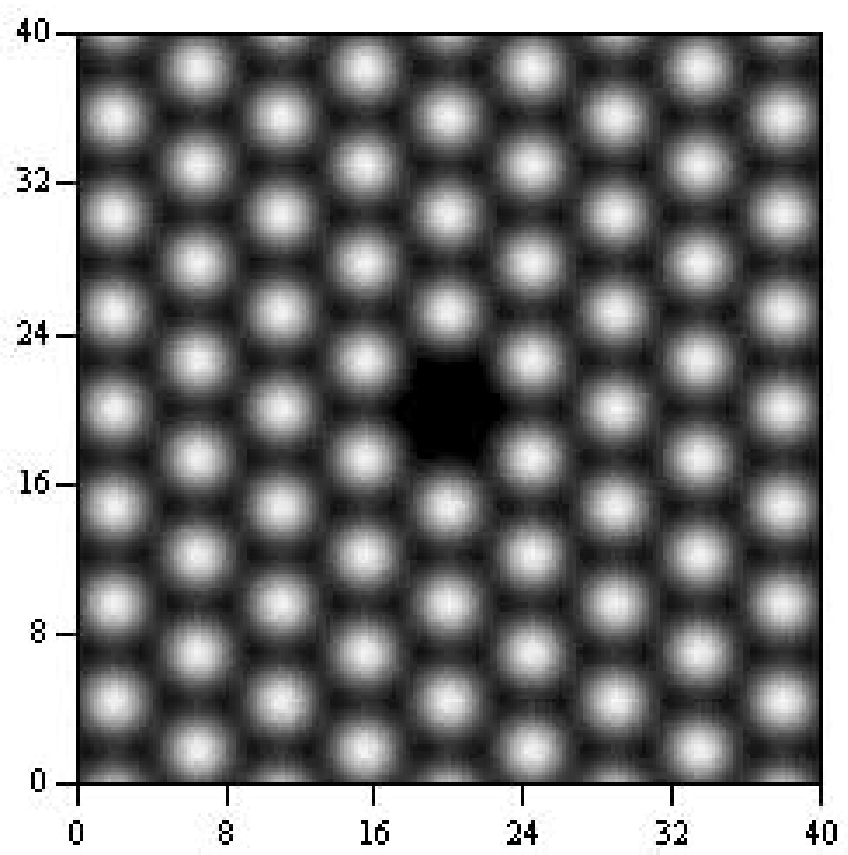}
b)\includegraphics[width= 0.3\textwidth]{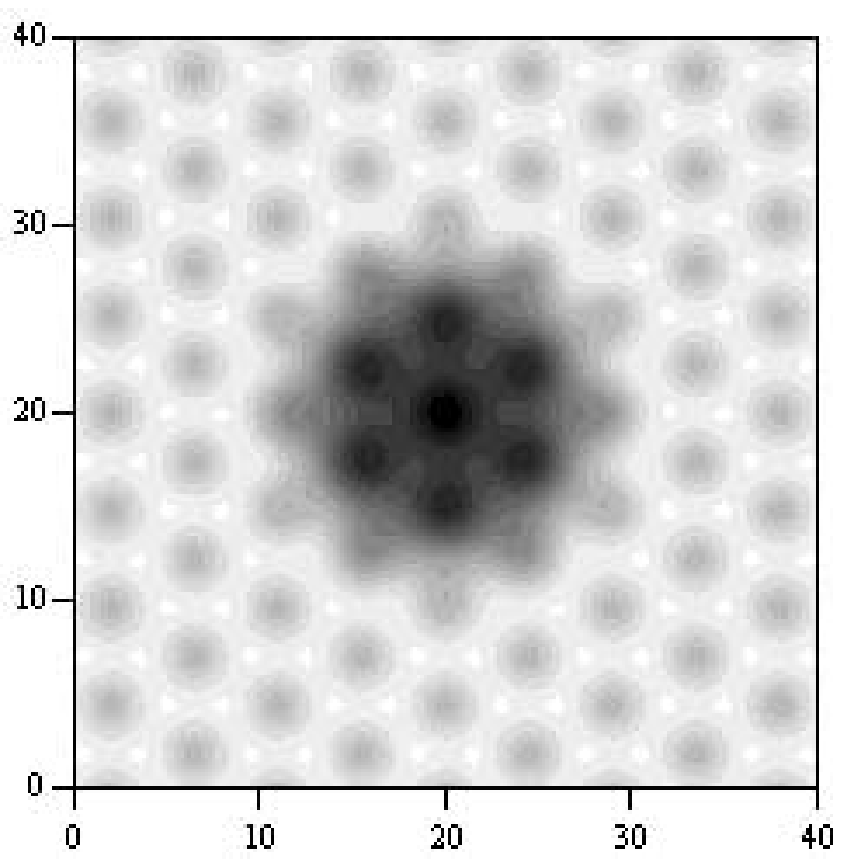}
c)\includegraphics[width= 0.3\textwidth]{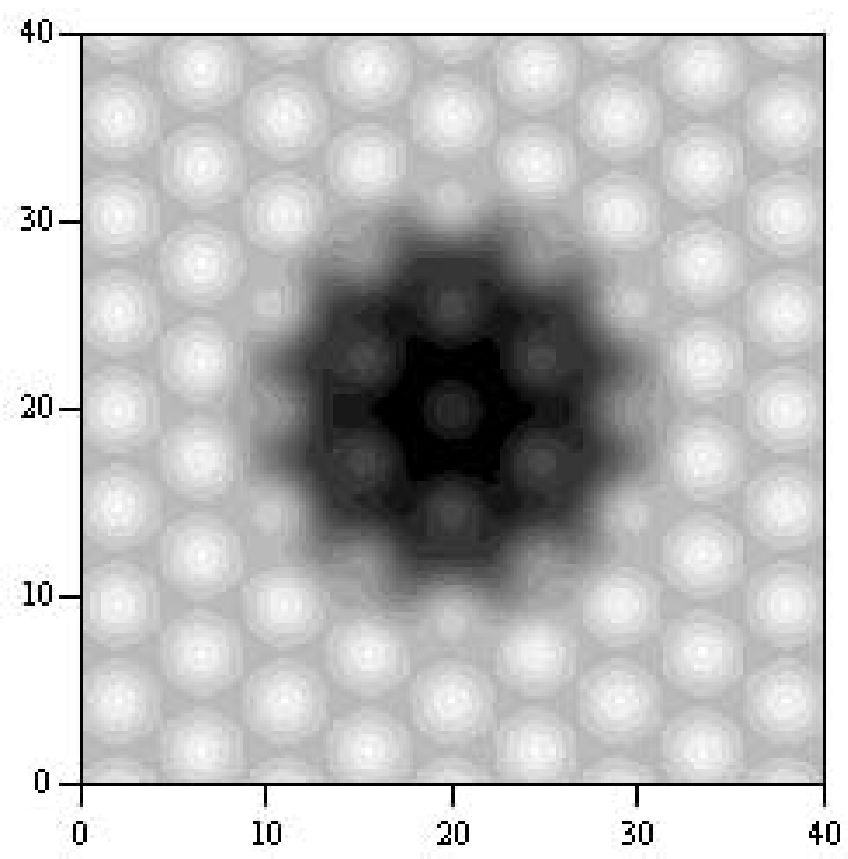}
d)\includegraphics[width= 0.3\textwidth]{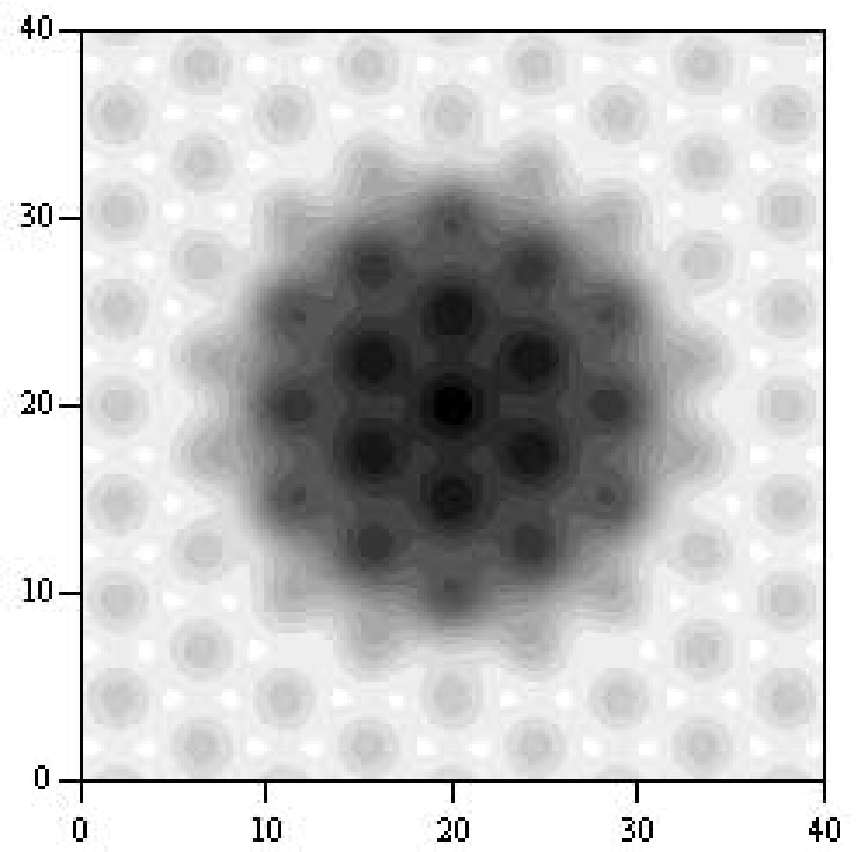}
e)\includegraphics[width= 0.3\textwidth]{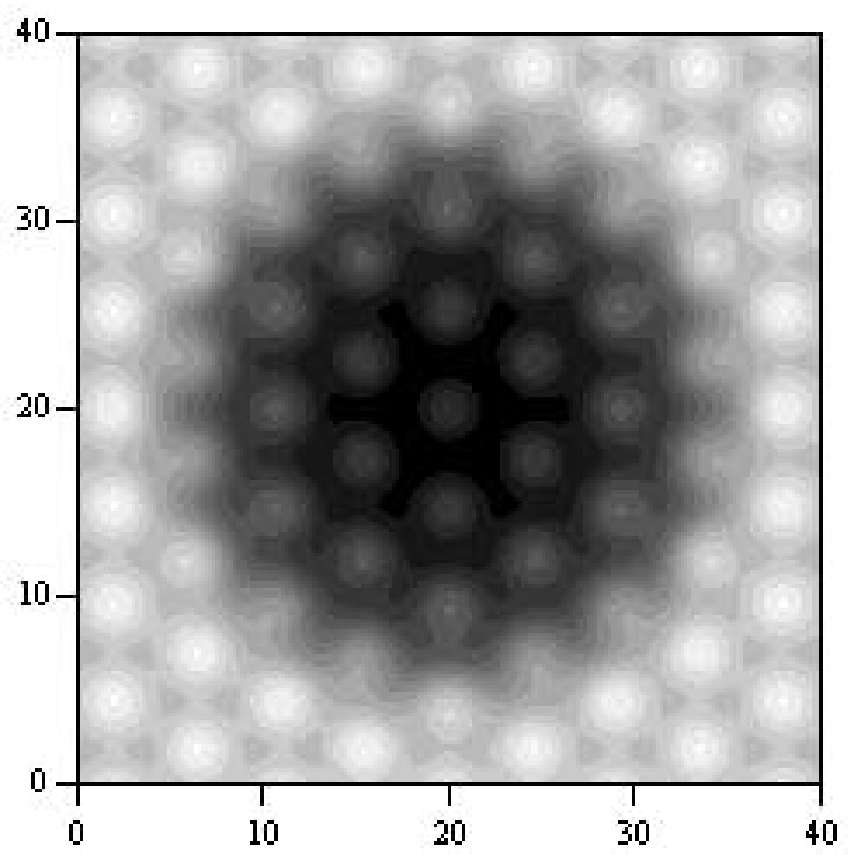}
f)\includegraphics[width= 0.3\textwidth]{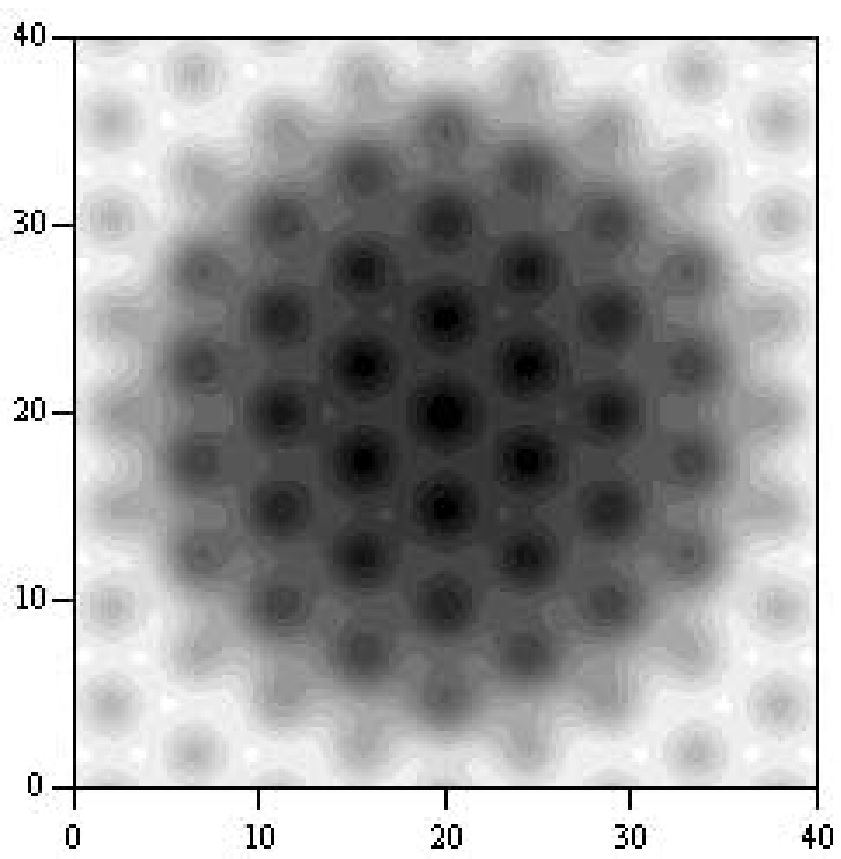}

\caption[]{Model function~(\ref{phi_1}) with single vacancy~(a) and AFM images calculated using Eq.\,(\ref{appar3}) for different values of contact area radius (0{,}9\,nm (b), 1{,}1\,nm (c), 1{,}3\,nm (d), 1{,}6\,nm (e) and 1{,}8\,nm (f)). \medskip \protect\\
{\footnotesize
The images show that point defect influence redistributes over area with size determined by $a$ value. The AFM images (b), (d) and (f) demonstrate inversion of contrast 

}
}
\label{defect}
\end{center}
\end{figure}

We applied the same model to describe the AFM visualization of single atomic vacancy using model function $\varphi '(x,y)$:
\begin{equation}
\label{phi_1}
\varphi '(x,y)=\varphi(x,y)-6e^{-\frac{k^2}{6}}(x^2+y^2)
\end{equation}
It was found that a vacancy contributes on large area of AFM image with size determined by value $a$, see fig.\,\ref{defect}, and ``false'' atom in the place of vacancy could be observed. 

So, we explain the peculiarities of AFM image formation (contrast inversion, hiding of single vacancy) on the basis of contact Herz theory.

\end{document}